\documentclass[10pt,journal,final,finalsubmission,twocolumn]{IEEEtran}

\title{Centrality measures and thermodynamic formalism for complex networks}

\usepackage[french]{babel}
\usepackage[latin1]{inputenc}
\usepackage[T1]{fontenc}

\author{Jean-Charles Delvenne \thanks{J.-C. D.'s present address: Université catholique de Louvain, Applied Maths Department, 4 Avenue Lemaître, B-1348 Louvain-la-Neuve, jean-charles.delvenne@uclouvain.be}  and Anne-Sophie Libert \\
Namur Center for Complex Systems (NAXYS)\\
Facultés Universitaires Notre-Dame de la Paix\\
Department of Mathematics\\
8, rempart de la Vierge \\
jean-charles.delvenne@fundp.ac.be, anne-sophie.libert@fundp.ac.be
}

\date{}

\usepackage{array}

\usepackage{latexsym}
\usepackage{amssymb}
\usepackage{amsmath}
\usepackage[dvips]{graphicx}
\usepackage{color}

\newtheorem{defi}{Definition}

\begin{document}

\maketitle

\bibliographystyle{plain}

\begin{abstract}

In the study of small and large networks it is customary to perform a
simple random walk,
where the random walker jumps from one node to one of its neighbours with
uniform probability.
The properties of this random walk are intimately related to the
combinatorial properties of the network.
In this paper we propose to use the Ruelle-Bowens random walk instead,
whose probability transitions are chosen in order to maximise the entropy
rate
of the walk on an unweighted graph. If the graph is weighted, then a free
energy is optimised instead of entropy rate.

Specifically, we introduce a centrality measure for large networks, which
is the stationary  distribution attained by the the Ruelle-Bowens random
walk; we name it Entropy Rank. We introduce a more general version, able
to deal with disconnected networks, under the name of Free Energy Rank. We
compare the properties of those centrality measures with the classic
PageRank and HITS on both toy and real-life examples, in particular their robustness to small modifications of the network. It is observed that our
centrality measures have a better discriminating power than PageRank,
being able to distinguish clearly pages that PageRank holds for almost
equally interesting, and is more sensitive to the medium-scale details of the graph.
%
\end{abstract}


\section{Introduction}

In the last decade tremendous amount of data has been collected on how various agents interact with each other.
This can be people exchanging phone calls in sociology, web pages pointing to each other through hyperlinks, genes influencing the expression of other genes in
genetics, food webs in ecology, etc.  These large to huge graphs require new powerful methods of analysis in order to identify the key structures of the graph.
A particular problem retains our attention here: centrality measures.

One of the most prominent application of centrality measures is the Web search, where the most central, best connected pages through the network of hyperlinks are often the most relevant regarding their content. Google and other Web search engines attribute
to each page of the Web a `PageRank' score, which measures
how well-connected this page is with respect to other pages
\cite{BrinPage98}. More specifically, a page has a high PageRank if
pointed to by pages with a high PageRank. Kleinberg
\cite{kleinberg99} has proposed the HITS method, where a page is a
good `hub' on a topic if it points to good `authorities' on this
topic, and a page is a good authority if pointed at by good hubs.
Other variants have been proposed by several authors, let us mention
only Ding et al. \cite{Ding03}, who propose a framework generalizing
HITS and PageRank and  Akian et al. \cite{Ninove}, who use
thermodynamic concepts in a different way from us. Those methods are all variants of the earlier Eigenvector centrality \cite{Bonacich87}, which computes the dominant left eigenvector of the adjacency matrix as the centrality measure. Other centrality measures, such as betweenness and closeness, based on counting shortest paths between nodes, have been popular \cite{Newman01}. Although the Web now constitutes the most spectacular application of centrality measures, they were first used for social networks analysis \cite{Bonacich87} and have found many other applications, most recently in economic networks \cite{Schweitzer2009,Bech2009}.

In this paper we apply methods from Ruelle's thermodynamic formalism to the field of large graphs, and in particular we introduce \emph{Entropy Rank} and \emph{Free Energy Rank} methods, which
rank the nodes of a network. 

Let us consider a strongly connected graph. While PageRank is based on the simple
random walk, where a random walker jumps from one node to any of its $d$ out-neighbours with uniform probability $1/d$, Entropy Rank is based on Ruelle-Bowens random walk \cite{Parry1964,ruelle78}. This random walk on the graph obeys transition probabilities wich are chosen in order to make all paths of same length to occur with approximately equal probability. In other words, the transition probabilities of the Ruelle-Bowens random walk are chosen so as to maximise the entropy rate of the random walk. Entropy Rank is now defined as the stationary distribution of the Ruelle-Bowens random walk. 

If the graph is not strongly connected, the Entropy Rank will have undesired effects, or even will not be uniquely defined. In this case, we use a trick close to PageRank's `teleportation' trick. Given any network,
one may complete the graph with all the non-edges and assign them a certain constant weight. We now have a weighted complete graph, with two different values for the edges. The Ruelle-Bowens random walk is also defined for weighted graphs, where the weights are interpreted as energies. Instead of maximising the entropy rate of the random walk, we maximise the sum of the entropy rate with the average energy of the edge; this sum is called the free energy of the random walk. As a result, the random walk will have a tendency to visit more often high-energy edges (it should be noted that Ruelle's sign convention for energy, which we follow here, is opposite to most physicists, who usually consider low energy to be more probable). The stationary distribution of the Ruelle-Bowens random walk on the complete weighted graph is what we call the Free Energy Rank.  

In the undirected case, they essentially coincide with Eigenvector centrality.
In the directed case, they share attributes with PageRank and HITS. For example, it attributes high scores to nodes which point to high-score nodes or are pointed to by high-score nodes, while only the latter is of direct relevance for the PageRank. 

We check on toy example and a 289K-node piece of the Web the ability of Free Energy Rank to better discriminate between the nodes. On the toy example, we notice that nodes identically ranked by PageRank are distinguished as different by Entropy and Free Energy Rank. In the large size example, we notice that the distribution of centrality scores is more inegalitarian for Free Energy Rank than for PageRank. It is thus better at separating `central' from `uncentral' nodes. Moreover, we introduce cliques, all nodes of which point to a single page, in order to see how the ranking of this page is enhanced. We observe that Free Energy Rank is more sensitive to such perturbations than PageRank.

The goal of this paper is therefore to introduce new centrality measures, and more generally to illustrate and promote the use of the Ruelle-Bowens random walk for complex networks. Although Ruelle's thermodynamic formalism, based on various powerful generalisations of Ruelle-Bowens random walks, is a physics-inspired, mathematically profound theory, it has received no attention so far from the community of large graphs and complex networks. Many algorithms proceed by performing a simple random walk on the graph in order to extract some combinatorial features. It has been shown in the area of community detection that different variants of the simple random walk (e.g., discrete-time or continuous-time) are able to highlight different features \cite{DelvenneYalirakiBarahona10, LambiotteDelvenneBarahona}.


\section{A reminder on PageRank}

\subsection{PageRank: First approach} 
\label{subsectpager1}

Let us now recall the principle of PageRank. PageRank can be defined in any kind network, as mentionned in the Introduction. Nevertheless, we will take as an explanatory example the case of the Web graph, with pages as nodes and hyperlinks as edges. Imagine a surfer starting from a page, and clicking randomly on the  hyperlinks on the page, each with equal probability. Repeating this process indefinitely, one may compute the asymptotic stationary probability distribution of the surfer. By elementary Markov chain theory, this distribution exists and does not depend on the initial state if the graph is strongly connected and aperiodic. It is given by the dominant left eigenvector of the row-stochastic, normalised adjacency matrix of the graph $D^{-1}A$. Here the adjacency matrix $A$ is defined by $A_{ij}=1$ if there is an edge from $i$ to $j$ and $A_{ij}=0$ otherwise, and $D$ is the diagonal matrix of outdegrees. The distribution, in the strongly connected aperiodic case, is also the vector of frequencies at which every node is visited by the random surfer. The PageRank \cite{BrinPage98} is then defined  as this stationary distribution.

The problem with this definition is that many graphs of interest, including the Web graph, are not strongly connected. In particular, many pages contain no hyperlink or are the target of no hyperlink. 
An improvement is therefore needed.

\subsection{PageRank with teleportation}
\label{subsectpager2}

To overcome this problem,  the possibility is given to the random surfer, with some probability $0< 1-\alpha
<1$, to jump to any other page of the web (with uniform
distribution). The surfer follows a hyperlink of the current page with
probability $\alpha$. If there is no hyperlink, then the surfer jumps to a random page with probability $1$ (we may call a teleportation, as this jump is not local).

Let $\tilde{A}$ be the adjacency matrix of the graph, with every
non-zero row normalised to $1$.
Then the stochastic matrix $M$ describing the Markov chain is
constructed as follows. Let $e$ be the vector of all ones,
normalised in order to sum to one. The $i$th row is equal to $(1-\alpha)
e^T
+ \alpha \tilde{A_i}$ if $\tilde{A_i}$  (the $i$th row of
$\tilde{A}$) is non-zero. If $\tilde{A_i}=0$ then the $i$th row is
taken as $e^T$.  The left dominant eigenvector of this matrix $M$,
normalised in order to sum to one, gives the unique stationary
distribution on the vertices. The PageRank is now defined as this stationary
distribution. 
Note that in practice, the entries of $e$ are not necessarily all equal but
can be chosen in order to favour some pages.

If $\alpha$ tends towards $1$, then we recover the first approach above (provided that the graph is aperiodic and strongly connected).
If $\alpha$ tends towards $0$, then the stationary distribution
tends towards the uniform distribution. For all $\alpha <1$, the PageRank is well defined on all graphs.  

\begin{figure}
\begin{center}
\includegraphics[bb=-10 -10 615 390,width=9cm,clip]{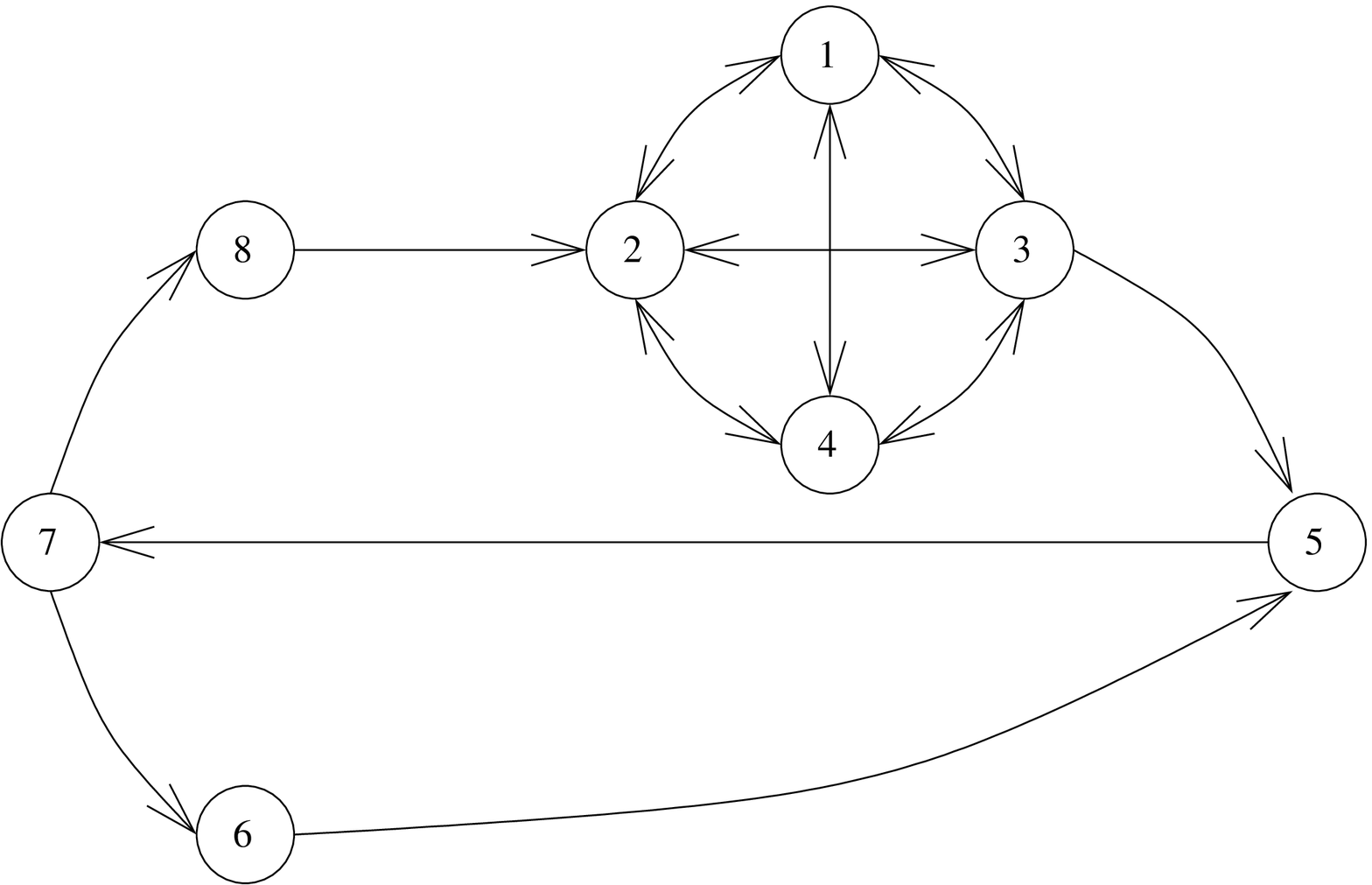}

\begin{tabular}{|c|c|c|c|c|}
  \hline
 \scriptsize{Vertex} & \scriptsize{PageRank} & \scriptsize{PageRank} & \scriptsize{Entropy Rank}
 & \scriptsize{Free Energy Rank} \\
   &  ($\alpha=1$)  &  ($\alpha=0.9$)  &   &  ($E=0.03$) \\
  \hline
 1& 0.1705 & 0.1549 & 0.2464 & 0.2400 \\
 2& 0.2045 & 0.1965 & 0.2487 & 0.2458\\
  3& 0.1818& 0.1644 & 0.2487 & 0.2460\\
 4& 0.1705 & 0.1549 & 0.2464 & 0.2400 \\
 5& 0.0909& 0.1035 &  0.0032 & 0.0099 \\
 6& 0.0455& 0.0601 & 0.0001 & 0.0019 \\
 7& 0.0909& 0.1057 & 0.0032 & 0.0076\\
 8& 0.0455&  0.0601 & 0.0031 & 0.0087\\
  \hline
\end{tabular}
\end{center}

\caption{Ranking scores according to different methods are computed
on this graph. Vertices 6 and 8 have the same PageRank, whatever value of
$\alpha$ is chosen, while both Entropy Rank and Free Energy Rank are able
to distinguish them. The gap of Entropy Rank between the best vertices
and worst vertices is larger than for any other method.}
    \label{fig_graph}
\end{figure}

PageRank has demonstrated its power in applications, on the Web and elsewhere. However we might argue that it may fail to distinguish the most interesting nodes in some cases. Indeed, let us take the graph of Figure
\ref{fig_graph}. Vertices $1$, $2$, $3$ and $4$ form a complete
directed subgraph, hence they concentrate most of the probability,
for values of $\alpha$ close to one, as expected. But they attribute an
equal probability to $6$ and $8$, as we can get easily convinced.
This might be argued as intuitively undesirable, because $8$ is
obviously a better page than $6$: it directly points to the most
interesting pages.

\section{Entropy Rank}
\label{sectentropy}

We now introduce a centrality measure that we call the \emph{Entropy Rank}. Assume again a surfer on a strongly connected, aperiodic graph. Instead of choosing a hyperlink with equal probabilities $1/d$, it chooses the first hyperlink on the page with a specific probability $p_1>0$, the second with a probability $p_2>0$, etc. We want to choose those probabilities in order to make the long term behaviour of the surfer as unpredictable as possible; in other words we want all the possible paths of the surfer (almost) equally probable.
  
Let us be more specific. Assume that on every page $i$ we have chosen probability of transition $p_{ij} >0$ towards page $j$ if there is a hyperlink from $i$ to $j$, and with probability $p_{ij}=0$ if there is none. This will result in the surfer being asymptotically in every state $i$ with a certain stationary probability $\pi_i$. The vector $\pi$ is the dominant left eigenvector of the row-stochastic matrix $M=(p_{ij})_{ij}$.
We may then compute the probability of the random surfer following the path $ijk \ldots mn$ as $\pi_i p_{ij} p_{jk} \ldots p_{mn}$. For every $t$, we may define the Shannon entropy $H(t)$ of all paths of length $t$ that the random surfer can follow. Then the \emph{entropy rate} of the random surfer is defined as $\limsup_{t \to \infty} \frac{H(t)}{t}$ (see, e.g., \cite{brown}) This entropy rate depends of course on the transition probabilities $p_{ij}$. We now want to choose the entries $p_{ij}$ in order to maximise the entropy rate of the random surfer. Let us see how to compute the $p_{ij}$ and the resulting entropy from the adjacency matrix $A$ of the graph.

The Shannon entropy of a probability distribution over a set of $N$
elements is at most $\log N$, and the uniform distribution is the
only distribution to achieve this bound, as well known.
Now consider a probability distribution of a random variable $X$ that is uniform up to a factor of
$a$, meaning that the probability of any event is at most
 $a/N$. Then the Shannon entropy of this
distribution is the convex combination of terms $-\log \textrm{Prob}(X=i)$,
every of which is
at least $\log N - \log a$. Hence the Shannon entropy itself is
at least $\log N - \log a$.

For any probability distribution over the paths, the Shannon entropy
of paths of length $t$ is at most $\log \#\{\textrm{paths of length
$t$}\}$. Hence the entropy rate is at most  $\limsup_{t \rightarrow
\infty} \frac{\log \#\{\textrm{paths of length $t$}\}}{t}$. This
last quantity is called the \emph{topological entropy} of the graph,
because it is not dependent on any particular probability
distribution, but is intrinsic to the graph. Since the number of paths
of length $t$ is the sum of all entries of $A^t$, the topological
entropy is readily seen to be equal to the logarithm of the spectral
radius of the adjacency matrix $A$.

Now, following Parry \cite{Parry1964}, we exhibit a particular
probability distribution whose entropy rate is precisely the topological entropy of the graph. 
Let $\lambda$ be the dominant eigenvalue of $A$ of maximal magnitude, $u$ be a nonnegative
left eigenvector for  $\lambda$ and $v$ be a nonnegative right
eigenvector for  $\lambda$. We thus have $u^TA=\lambda u^T$ and $Av=\lambda
v$. Their existence is ensured by
Perron-Frobenius theorem, and they can be computed by the power method. Normalise
$u$ such that $\sum_i u_i = 1$,
and normalise $v$ such that $\sum_i u_iv_i =1$. Choose the
probability $p_{ij}$ to take the existing edge $(i,j)$ starting
from $i$, to be $v_j/\lambda v_i$. This is indeed a probability
distribution over the outgoing edges of $i$, since $\sum_{j:(i,j)
\textrm{is an edge}}  v_j/ \lambda v_i=\lambda^{-1}(Av)_i/
v_i=1$.  Then the row-stochastic transition matrix
$M=\lambda^{-1} \textrm{diag}(v)^{-1} A\,\,\, \textrm{diag}(v)$, where
$\textrm{diag}(v)$ is the diagonal matrix formed from vector $v$.

The distribution attributing a probability $\pi_i=u_iv_i$ to vertex
$i$ is an invariant distribution on the vertices of the Markov chain. 
Indeed, $\pi^T  M= \pi^T  \lambda^{-1} \textrm{diag}(v)^{-1} A \quad
\textrm{diag}(v)=u^T \lambda^{-1} A \quad 
\textrm{diag}(v)=u^T\textrm{diag}(v)=\pi^T.$

 The probability of path $ij$ is $u_iv_i \lambda^{-1}v_j/ v_i= \lambda^{-1}u_i v_j$,
the probability of path $ijk$ is $\lambda^{-1}u_i v_j
\lambda^{-1} v_k/ v_j=\lambda^{-2}u_iv_k$, and more generally any
path of length $t$ going from vertex $i$ to vertex $j$ has a
probability $\lambda^{-t}u_iv_j$ (which does not depend on the intermediate
vertices). We know that the number of paths of length
$t$ is in the order of $\lambda^t$ (up to a factor).
Hence the probability distribution over paths of fixed length is uniform
up to a factor (which does not depend on $t$).
The Shannon entropy of paths of
length $t$ therefore grows as $t \log \lambda$, up to an additive constant.
The entropy rate of this distribution is thus $\log \lambda$, which
is optimal.

In brief, we have proved the following facts:

\begin{itemize}
\item the behaviour of a random surfer with maximal entropy rate can be computed
from a
left and right nonnegative dominant eigenvector, obtained for instance with the power
method;
\item the resulting distribution on vertices is given by the componentwise
product of the two eigenvectors;
\end{itemize}

\begin{defi}
The \emph{Entropy Rank} of vertex $i$ of an unweighted strongly connected aperiodic graph is defined as
the probability $u_iv_i$, where $u$ ($v$) is the left (right) dominant eigenvector
of the adjacency matrix.
\end{defi}

Since the graph is strongly connected and aperiodic, then $\lambda$, $u$
and $v$ are unique and positive, by Perron-Frobenius theorem.
The Entropy Rank is then uniquely defined and non-zero
on every vertex. Note that the matrix $\lambda^ {-t} A^t$ can be shown to converge to
$vu^T$, whose diagonal gives the vertex probability  distribution.
As shown in \cite{Parry1964}, when the graph is strongly
connected there is no other probability distribution that maximises
the entropy rate.  See a numerical example on Figure \ref{fig_graph}.
A more trivial example is the complete graph on $n$
vertices, for which $A$ is the matrix of ones (except on the diagonal); 
 we see that the entropy rate has the maximal value $\log(n-1)$ for 
the uniform distribution.

Note also that if we reverse all edges of the graph, then the matrix
$A$ is replaced by $A^T$, the vectors $u$ and $v$ switch their roles
and the final value for the Entropy Rank is the same. Hence the
entropy method takes into account, not only the paths leading to a
vertex, but also the paths issued from a vertex. In the case of an undirected graph, as
both eigenvectors are identical, Entropy Rank provides the same ranking as Eigenvector centrality, which ranks nodes according to their entry of the left eigenvector.

\section{Free Energy Rank}
\label{sectfreeenergy}

We want  a method giving to every graph, even non strongly connected, a unique centrality score of the nodes, 
which is non-zero on every vertex. That is why we add the following
improvement, which is a particular case of Ruelle's thermodynamic
formalism \cite{ruelle78}. On the complete directed graph with self-loops
that extends the original graph 
we attribute an `energy' $U=0$ to the edges of the original graph
and an `energy' $U=-U_0 <0$ to the other edges. Now consider
the set of all paths in the complete graph. The energy of a path is defined as
the energy of its first edge. On this set we want to put an
invariant probability measure that maximises the quantity
$S+\overline{U}$, where $S$ is the entropy rate and $\overline{U}$ is the
expected energy for the probability measure. The maximum of this
quantity is analogous to what is called `free energy' in
thermodynamics (with unit temperature and up to the sign). It is
also called `topological pressure' in the literature of
thermodynamic formalism.

This time we consider the matrix $B$ such that $B_{ij}=
\exp(U_{ij})$, where $U_{ij}$ is the energy of the edge $ij$. Note
that if $U_0 \rightarrow \infty$, then $B$ converges to the
adjacency matrix $A$. Note also that the matrix $B$ can be obtained
from $A$ by replacing zero entries with $e^{-U_0}$.

It is possible to see that the maximising set of transition probabilities 
exists and is unique, and we can compute it in the following way.
Let $\lambda, u,v$ be such that

\begin{itemize}
    \item $\lambda$ is the dominant eigenvalue of $B$;
    \item $u^TB=\lambda u^T$ (left eigenvector);
    \item $Bv=\lambda v$ (right eigenvector);
    \item $u > 0$, $\sum_i u_i=1$;  
    \item $v > 0$, $\sum u_i v_i=1$;  
\end{itemize}

These objects exist and are unique, by Perron-Frobenius theorem.

Now, we claim that the maximising transition probabilities gives a probability of $u_i v_i$
to be
in vertex $i$. It also gives
a probability of $\lambda^{-1} U_{ij} v_j/v_i$ for the transition $i \rightarrow
j$, of energy $U_{ij}$. These claims can be derived as corollaries
 to Ruelle's more general results \cite{ruelle78}, but we prefer to give
 an elementary argument for the sake of self-containedness.
\begin{defi}
For a given $\epsilon>0$, the \emph{Free Energy Rank} of vertex $i$ of an
unweighted directed graph  is defined as the probability $u_iv_i$, where
$u$ ($v$) is the left (right)
 dominant eigenvector of the matrix $B$ obtained from the adjacency matrix
 by replacing the zero entries with $e^{-\epsilon}$.
\end{defi}

The proof of the claim, which we give for the sake of clarity, relies on
the  following result, well known in statistical physics; see for instance
\cite{ruelle78}. Given a finite set
endowed with a real-valued energy function, the only probability distribution on this set
that maximises the \emph{free energy} (sum of Shannon entropy and expected
energy) is the \emph{Boltzmann distribution}, attributing probability
$\exp(U_i)/\sum_i \exp(U_i)$ to element $i$. The free energy is then
$\log \sum_i \exp(U_i)$. If a probability distribution is the Boltzmann
distribution up to a factor $a$, meaning that the probability
for element $i$ is
at most $a \exp(U_i)/\sum_i \exp(U_i)$, then the
corresponding free energy is at least $\log \sum_i \exp(U_i)-\log
a$.

The random walk described just above gives a probability
$\lambda^{-t} \exp(\sum U_{kl}) u_i v_j $ to a path of length $t$
from vertex $i$ to vertex $j$, where $\sum U_{kl}$ is the sum of
energies of all edges $kl$ on the path. This has the form of a
Boltzmann distribution, up to a factor. Now if we give to a path of
length $t$ a  `path energy' that is the sum of all energies of its $t$
individual edges, then this probability distribution  yields a
`path free energy'  equal to $\log \sum_{\textrm{paths of length
$t$}} \exp(\sum U_{kl})$ up to an additive constant (independent of
$t$), which is almost maximal. This path free energy, divided by $t$,
gives for $t \to \infty$ a maximal  $S+\overline{U}$. Note that the
expected energy of a path of length $t$ is exactly $t \overline{U}$,
Note also that the maximal free energy is again  $\log \lambda$, the
logarithm of spectral radius of $B$.

The interpretation of this framework is the following: a random
surfer can jump from any page to any page, with an energy cost of $U_0$
if no hyperlink is present between the pages.
The surfer, whose aim is to optimise the free energy $S+\overline{U}$,  is
therefore incited to follow
hyperlinks (edges of the graph) in priority. If the energy gap $U_0$
is $0$, then the optimal probability is uniform. If the energy gap is
high, then the surfer
is incited to follow hyperlinks most of the time. Such a phenomenon
is similar to what is observed when varying the factor $\alpha$
between $0$ in $1$ in the PageRank method (as detailed in Section
\ref{subsectpager2}).

One may ask how to choose a reasonable $U_0$. Knowing that $\alpha=0.85$ or $\alpha=0.9$, for instance, works well in the case of PageRank, one may develop a heuristic argument to find a corresponding value of $U_0$ as follows. Suppose that the outdegree of the graph is constant. Then the right eigenvector of $A$ is constant as well, and the PageRank (for $\alpha=1$) is equal to the Entropy Rank, if defined. Moreover, for every value of $\alpha$, there is a corresponding value of $U_0$ such that the PageRank and the Free Energy Rank coincide. A calculation shows that this value is such that $E=e^{-U_0}=1/(1+\frac{\alpha N}{(1-\alpha)d})$.
When the graph is not with constant ourdegree, PageRank and EntropyRank do not coincide in general. However we may take $d$ as the average outdegree to guess a reasonable value for $E$. 

The free energy method also gives a non-zero
probability to any vertex of the graph. An example of calculation is
shown on Figure \ref{fig_graph}. We consider a value of $U_0$ equivalent to $\alpha=0.9$. Let us remark that the node 7 has this time a Free Energy Rank lower than the node 8, which indicates that the page 8 is more interesting, which is a sensible claim. 

Again, the Free Energy Rank is invariant under reversal of edges.

\section{Numerical experiments}

\begin{figure}
\includegraphics[height=7cm]{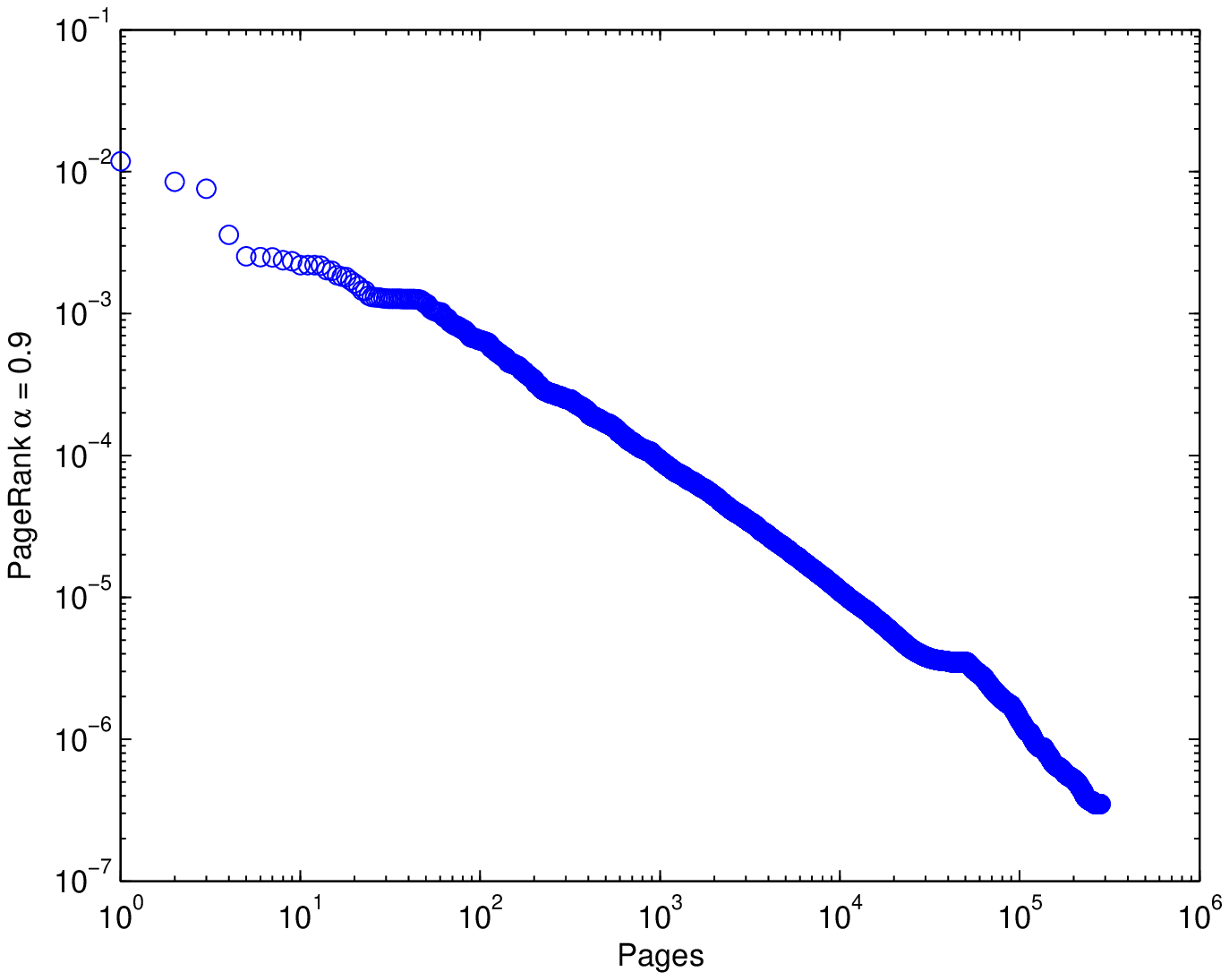}
\includegraphics[height=7cm]{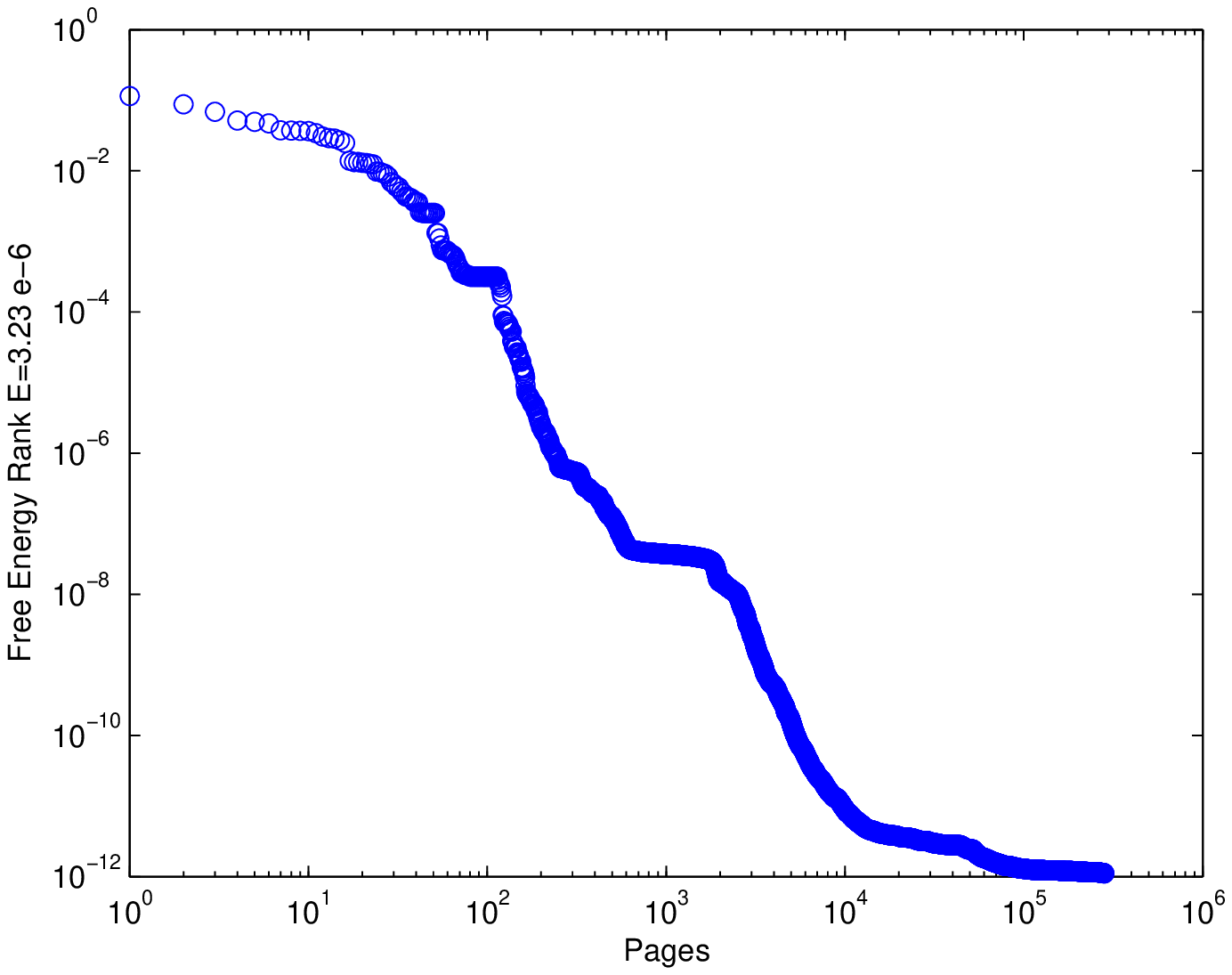}
\includegraphics[height=7cm]{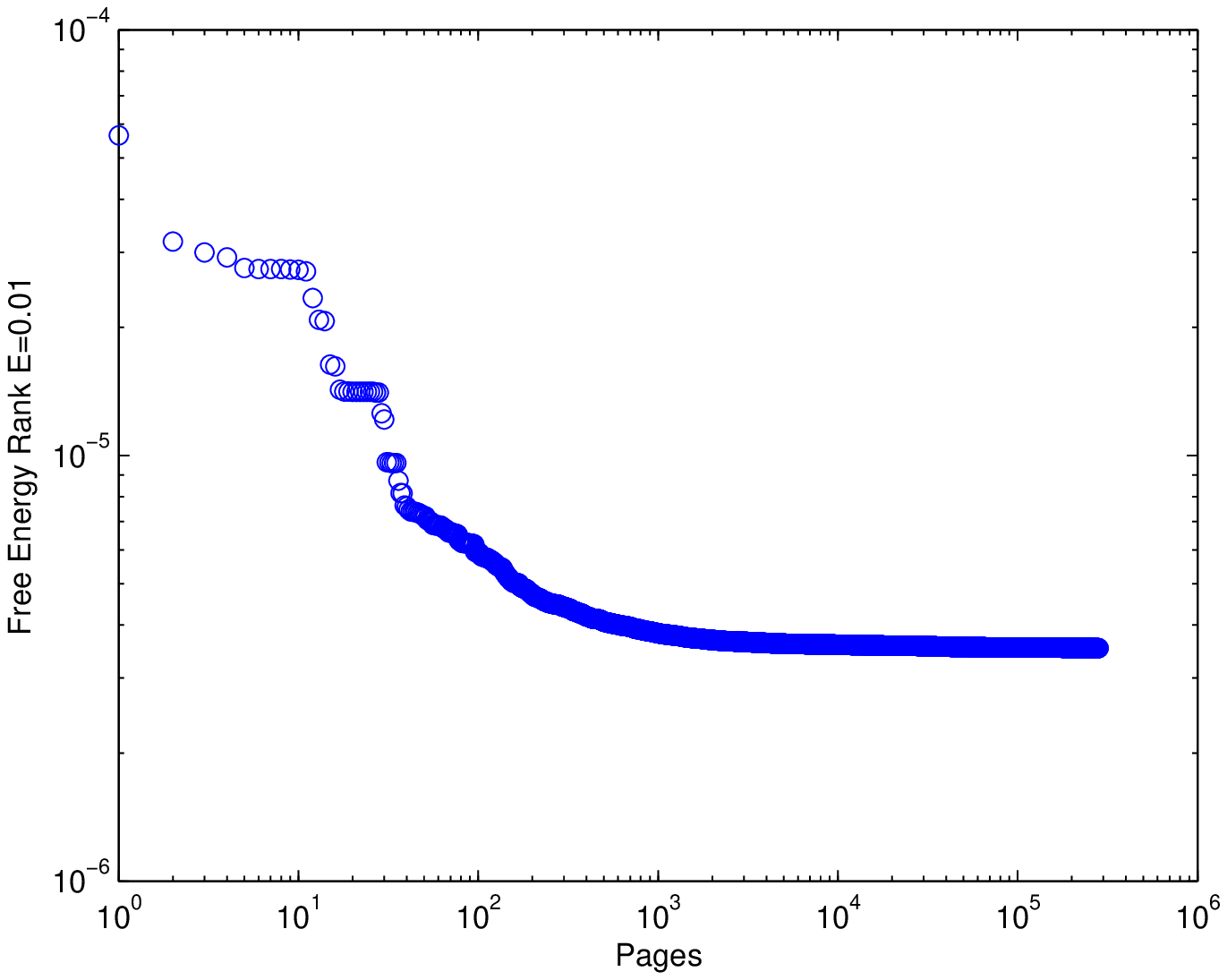}
\caption{Pagerank and Free Energy Rank distributions in logarithmic scales. \emph{Top:} The PageRank seems to be distributed according to a power law , of slope close to $-1$. \emph{Center:} The distribution curve of Free Energy Rank is steeper, which indicates a larger discriminating power between the `best' and `worst'  pages. Here $E=3.23 \, 10^{-6}$, which corresponds to $\alpha=0.9$ for PageRank. The distribution is also less regular than for above. \emph{Bottom:} A larger value of $E=0.01$ limits the spread of the distribution and creates an almost uniform distribution for the `worst' pages.}
\label{distributions}
\end{figure}

We now compare the distribution of PageRank and Free Energy Rank for a 289000-node piece of the Stanford Web \cite{kamvar}. The distributions are shown in Figure \ref{distributions}. The first panel indicates the Pagerank scores associated to $\alpha=0.9$ and shows the well-known power law trend of the Pagerank distribution. The Free Energy Rank distribution with an equivalent value of $U_0$ is represented on the second panel. While the qualitative behaviour is similar, the values of Free Energy Rank are more spread out: the ratio of centrality score between the best and worst pages is much higher when the centrality score is Free Energy Rank rather than PageRank. Finally, for smaller value of $U_0$ (or equivalently larger value of $E$), the main pages are highlighted in the distribution (lower panel of Figure \ref{distributions}), while the worst pages are gathered to the same Free Energy Rank value. This is not surprising, since the distribution is determined by the left and right eigenvector of a matrix $B$ whose entries are all $1$ or $E$. 
Therefore, the right eigenvector has entries whose ratio is at most $1/E$, and similarly for the left eigenvector. Thus, the ratio between two probabilities is at most $1/E^2$, which limits the spread between the `best' and `worst' pages.


Therefore a high value of $E$ is interesting in some circumstances, when we only want to distinguish the good pages between them, and leave all the bad pages to virtually the same value. 

\section{The effect of link farms}

Intuitively, a node has a high Entropy Rank/Free Energy Rank if it belongs to many paths. Thus there are two ways to get a high Entropy Rank or Free Energy
Rank: to be pointed by good pages or to point to good pages. This is
reminiscent of HITS method \cite{kleinberg99}, that computes a hub
score and authority score for every node from the dominant
eigenvectors of $AA^T$ and $A^TA$. The exact relationship between
HITS and the entropy method remains to be investigated. Let us now see how easy it is for a malicious webmaster to artificially boost its ranking by creating a link farm, i.e. a large group of dummy pages whose structure is designed to improve the ranking of a specific page.

A typical way to increase the PageRank score of a page consists in changing the page into a good authority, by adding a large number of pages all pointing to each other and pointing to the page to be artificially increased. This technique has an interesting impact on the Free Energy Rank score, as shown by the following simulation. 

For the piece of the Stanford Web used in the previous section, we choose the page which was classified at rank 200000th according to Free Energy Rank, for $E=3.23 e-6$, and classified at rank 154325th according to PageRank for $\alpha=0.9$. We then added a link farm of a hundred nodes pointing to each other and to this page. This page then reached rank 627th according to Free Energy ranking and rank 29173th according to PageRank. Interestingly, the hundred new pages get an even (slightly) higher Free Energy ranking than the page they are conspiring to push forward, while they get a much lower ranking than this same page for the PageRank.

\begin{figure}
\includegraphics[height=7cm]{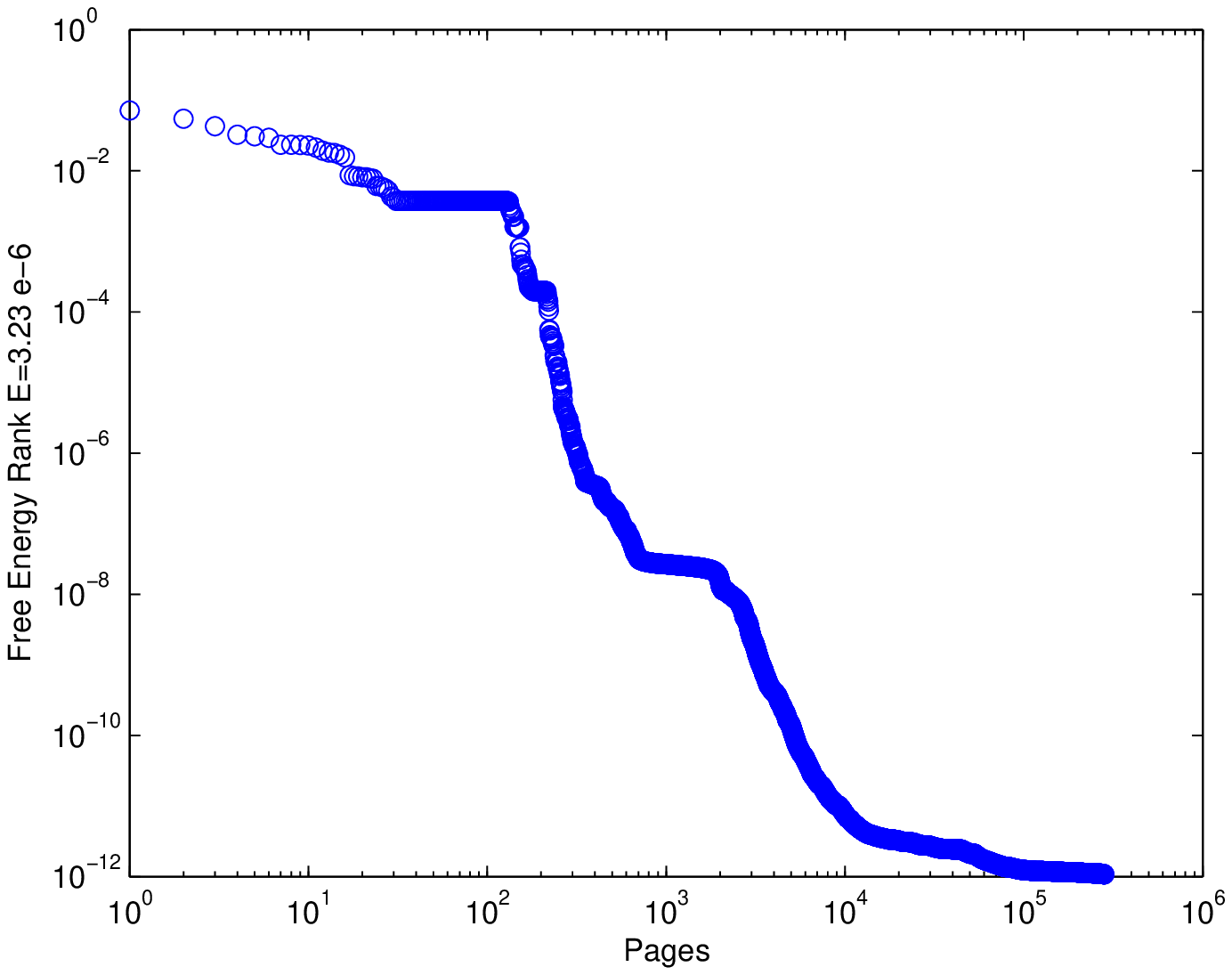}
\caption{Sensitivity of the Free Energy Rank distribution of Fig. \ref{distributions} ($E=3.23 10^{-6}$) to a link farm of a hundred nodes pointing to each other and to the page initially classified at rank 200000th. Notice, comparing with Figure \ref{distributions}(center),  how the hundred nodes of the link farm receive a very high ranking, thus forming a new plateau in the Free Energy Rank distribution.}
\label{tricherie}
\end{figure}

Although the rank benefit is larger for the Free Energy Rank method, the cheating is thus easier to detect: a new plateau has appeared in the distribution of centrality around ranks 30th-130th; see Figure \ref{tricherie}.  Since the nodes in the link farm do not get as much PageRank as the page they push forward, this plateau exists, but is much less visible in the distribution curve of PageRank (around ranks 54180th-54280th). 

In the Hits method, we know that pointing to good pages on a topic increases the hub score of a page. Again we verify the impact of this falsification on Free Energy Rank. We choose again the same 200000th page according to Free Energy Rank with $E=3.23 \, e-6$. We make it point to the 100th best pages. As a result, the 200000th page is now 629th. 

Note than in the HITS method, a page largely improves its authority score  when pointed to by a link farm of a hundred pages, and largely improves its hub score when pointing to the best pages, but this improvement is still much less impressive than for the corresponding falsification of the Free Energy Rank.


\section{Extensions}

Note that the experiences above refer to the application of centrality measures for the full graph of the Web. Those centrality measures can be applied in a number of contexts. For example, in the HITS method it is usually considered that scores are computed only on the subgraph composed of those pages which contain a certain keyword and their neighbours. In this example and in other kinds of networks, such as the interbank network \cite{Bech2009}, where nodes are banks and edges loans between them, the techniques of cheating  do not make sense of course, or not in the same way. More generally, the meaning of the different centrality measures vary according to the meaning of the network, and according to the example one may consider one or another centrality measure to be more or less appropriate in such or such network.

We have applied the Free Energy Rank by endowing every non-edge $ij$ with a certain energy $U_0$. Of course we could choose a non uniform energy, which would depend on $i,j$ or both. This is similar to Google's replacing in PageRank the jump uniformly to any other node by a jump to any other node, with a probability that depends on this node, in order to favour some nodes.

So far we have consider unweighted networks. On a weighted network, we can interpret the weights as energies (or even exponential if those weights are nonnegative) and define a centrality measure from the stationary distribution maximising the free energy. If the graph is not strongly connected and aperiodic, then we can use the same trick again of transforming every non-edge into an edge with a certain energy $U_0$.

The Ruelle-Bowens random walk is not defined locally; instead, its transitions depend on the whole graph. Approximations of the Ruelle-Bowens random walk that can be performed with local information are explored in \cite{SinatraLambiotteetal}.

\section{Conclusions}

We showed how to use the Ruelle-Bowens free-energy maximising random walk on
any weighted graph instead of the simple random walk in order to extract information from this graph.

We applied it to centrality measures, introducing Entropy Rank and Free Energy Rank, comparing it with PageRank and HITS. We compared the robustness of those centrality measures with respect to the introduction of a clique, called a `link farm' when it comes to fraudulent manipulation of the Wep page rankings. We observed that the Free Energy Rank is much more sensitive to such perturbations than the PageRank and HITS. This suggests that global distribution of Free Energy Rank is more sensitive to the medium scale details of the graph, and may explain why it does not appear as power-like as PageRank does.

We do not claim that Ruelle-Bowens provides a better basis for centrality, only that it provides a completely different spectral centrality than considered so far, with very different properties, which may be more or less suitable according to the context.

We insist that this is one possible application of the Ruelle-Bowens random walk to complex networks. Every method that performs a random walk on the graph in order to analyse it, such as Markov clustering \cite{vanDongen00}, Walk Trap \cite{LatapyPons08}, stability \cite{DelvenneYalirakiBarahona10, LambiotteDelvenneBarahona}, commute-time distance \cite{FoussPirotteRendersSaerens06}, kWalks \cite{Dupont06}, etc., could in principle be adapted to the Ruelle-Bowens walk. Again, the resulting algorithms would perhaps be more relevant in some cases and less so in others. The exploration of such algorithms and for which applications they are suitable opens a vast field which we leave for future research.

%
%
%
%

\section{Acknowledgements}
The work of A.-S. L. is supported by an FNRS post-doctoral research
fellowship. The work of  J.-C. D. is supported by the Belgian Programme on Interuniversity Attraction Poles initiated by the Belgian Federal Science Policy Office, and by the Concerted Research Action (ARC) ``Large Graphs and Networks'' of the French Community of Belgium. The scientific responsibility rests with its authors. This paper has benefitted from discussions with the team
GYROWEB at INRIA, Rocquencourt (France) and the team Large Graphs
And Networks at Universit\'{e} catholique de Louvain, Louvain-la-Neuve
(Belgium). In particular, Diep Ho Ngoc simplified the main proof in
Section \ref{sectentropy}.




\end{document}